\documentclass[showpacs,pra,aps]{revtex4}
\usepackage{graphicx}
\def\be{\begin{equation}}
\def\ee{\end{equation}}
\def\bea{\begin{eqnarray}}
\def\eea{\end{eqnarray}}
\def\bfr{{\bf r}}
\def\bfq{{\bf q}}
\def\bfs{{\bf s}}
\begin{document}
\title{Some exact results for a trapped quantum
gas at finite temperature}
\author{Brandon P. van Zyl, Rajat K. Bhaduri}
\affiliation{Department of Physics and Astronomy,
McMaster University, Hamilton,
Ontario, Canada, L8S~4M1}
\author{Akira Suzuki}
\affiliation{Tokyo University of Science, Kagurazaka 1-3, Shinjuku-ku, 
Tokyo 162-8601, Japan}
\author{Matthias Brack}
\affiliation{Institut fu\"r Theoretische Physik, Universit\"at
Regensburg, D-93040 Regensburg, Germany}

\begin{abstract}
We present closed analytical expressions for the particle and kinetic 
energy spatial densities at finite temperatures for a system of 
noninteracting fermions (bosons) trapped in a $d$-dimensional harmonic 
oscillator potential. 
For $d=2$ and $3$, exact expressions for the $N$-particle densities are used to 
calculate perturbatively the temperature dependence of the splittings of the 
energy levels in a given shell due to a very weak interparticle interaction 
in a dilute Fermi gas. In two dimensions, we obtain analytically 
the surprising result
that the $l-$degeneracy in a harmonic oscillator shell is {\it not} lifted 
in the lowest order even when the exact, rather than the Thomas-Fermi 
expression for the particle density is used.    
We also demonstrate rigorously (in two dimensions)
the reduction of the exact zero-temperature fermionic expressions to the
Thomas-Fermi form in the large-$N$ limit.
\end{abstract}

\maketitle
%%%%%%%%%%%%%%%%%%%%%%%%%%%%%%%%%%%%%%%%%%%%
\section{Introduction}
\label{intro}
The observation of Bose-Einstein condensation in ultra-low
temperature trapped atomic gases 
(see e.g., Ref.~\cite{bec} for a comprehensive review), along with the 
recent experimental effort of Jin and DeMarco \cite{jin} in 
achieving a degenerate Fermi gas,
has sparked a renewed interest in the study of the thermodynamic properties
of trapped quantum gases. 
For a dilute gas of trapped bosons or fermions, assuming the 
interparticle interactions to be very weak,
the many-body system may be modeled in the zeroth-order approximation 
as a system of noninteracting particles.  This is an
attractive situation from a theoretical point of view because the 
thermodynamics
of the noninteracting $N$-body system can be calculated from the single-particle
partition function of the associated trapping potential.
For the case of Bose statistics, exact closed form expressions for the
thermodynamic quantities of the ideal trapped gas can be obtained at all 
temperatures in {\em any} dimension provided the trapping potential is an
harmonic oscillator (HO).
An ideal trapped gas of fermions on the other hand is not as easy to analyze.
Unlike the Bose gas, the Fermi distribution function cannot be expanded as
a power series for all temperatures, and as a result, closed form expressions
for e.g., the particle and kinetic energy densities, at
finite temperature are difficult to obtain.  Indeed, it is 
only recently that Brack and van Zyl~\cite{brack2} have constructed exact, 
analytical expressions for the particle and kinetic energy densities of 
a harmonically confined ideal Fermi gas in any spatial
dimension.  These expressions are useful for both numerical 
and analytical computations, but are restricted to zero temperature.
Prior to the work in Ref.~\cite{brack2},
fermions in harmonic traps have been studied using exact {\em numerical} 
\cite{schneider,brosens,vig,akdeniz}, analytical \cite{bruun,gleisberg,wang} 
and semiclassical techniques \cite{butts}.  
The exact analytical results for the particle and kinetic energy densities,
however, are
limited to one dimension (1D) at zero temperature \cite{gleisberg} or
arbitrary dimensions in the high temperature regime \cite{wang,note1}.

In this article, our main emphasis will be on presenting exact 
analytical results 
for an ideal $d$-dimensional trapped Fermi gas at 
{\em any finite temperature}, thereby 
extending the work in Ref.~\cite{brack2} beyond $T=0$.  We shall also 
digress to give some novel analogous expressions for the ideal trapped Bose 
gas which have not yet appeared in the literature. 
The trapping 
potential will be taken to be an isotropic HO, although our 
results are easily generalized to the anisotropic case.  Apart from the 
intrinsic merit of exact results for a many-body system, our findings
should also prove useful in the density-functional theory
of weakly interacting inhomogeneous Fermi systems.
%are also useful in the perturbative treatment of a  
%very weakly interacting dilute Fermi gas.

The rest of our paper is organized as follows:  In Sec.~II, we derive
analytical expressions for the finite-temperature particle and kinetic
energy densities for an ideal trapped Fermi gas in arbitrary dimensions.
We also show rigorously how the exact quantum expressions at $T=0$ 
asymptotically approach their
Thomas-Fermi (TF) results in the large-$N$ limit; 
we know of 
only two other examples where such a result has been rigorously 
established analytically~\cite{gleisberg,lieb}. 
Then, in Sec.~III, we derive expressions analogous to the fermionic system
for a trapped gas of noninteracting bosons.  In Sec.~IV,
we apply our {\em exact} result for the spatial density to calculate the 
mean-field potential, and the resulting finite-temperature 
splitting of the energy levels in a $d=3$ HO shell perturbatively. 
Although a similar {\em zero temperature} calculation has already 
been considered numerically by Heiselberg and Mottelson~\cite{heiselberg},
our results (in contrast to Ref.~\cite{heiselberg})
do not rely on the TF approximation for the particle density. 
The corresponding calculation for $d=2$ yields an interesting result, which is
to be discussed in Sec.~IV B.  Finally, 
in Sec.~V, we summarize our results and suggest
other possible avenues for future research.
%%%%%%%%%%%%%%%%%%%%%%%%%%%%%
\section{Ideal Fermi gas at finite temperature}
\label{fermigas}
\subsection{Single-particle density in $d$ dimensions}
\label{fermidensity}
The starting point in our investigation is a system of noninteracting
fermions at zero temperature 
described by the time-independent Schr\"odinger equation
\be
{\hat H} \phi_i(\bfr) = [\hat{T} + V(\bfr)]\phi_i(\bfr) = \varepsilon_i
\phi_i(\bfr)~,
\ee
where $V(\bfr)$ is a one-body potential to be specified later (all 
$\epsilon_i$'s are taken to be positive).   
The single-particle density matrix can be obtained by an inverse 
Laplace transform
of the zero-temperature Bloch density matrix, $C_0(\bfr,\bfr')$:
\be
\rho(\bfr,\bfr') = 2 \sum_{\varepsilon_i < E_f}\phi_i^{\star}(\bfr')
\phi_i(\bfr) \Theta(E_f) = {\cal L}_{E_f}^{-1}\left[\frac{2}{\beta}C_0(\bfr,\bfr';\beta)
\right]~,
\label{onesided}
\ee
where 
\be
C_0(\bfr,\bfr';\beta) = \sum_{{\rm all~} i}\phi_i^{\star}(\bfr')\phi_i(\bfr)
\exp(-\beta\varepsilon_i)~,
\label{bloch0}
\ee
and $E_f$ is the Fermi energy; the factor of two accounts for spin. We have 
put in the unit step-function $\Theta(E_f)$ in Eq.~(\ref{onesided}) so that 
the Laplace transform with respect to $E_f$ may formally be taken to be 
two-sided~\cite{pol}.
Note that in quantum statistical mechanics, $\beta$ is usually identified with the
inverse temperature, $\beta = 1/k_B T$.
However, in our present context, $\beta$ is to be interpreted as mathematical 
variable which in general is taken to be complex, and {\it not} the inverse 
temperature $1/k_BT$.

At finite-temperature, the single-particle density matrix is 
obtained from
the Bloch density matrix by using the relation \cite{brack}
\be
\rho(\bfr,\bfr';T) = {\cal L}^{-1}_{\mu}\left[ \frac{2}{\beta}
C_T(\bfr,\bfr';\beta)\right]~,
\label{rhoT}
\ee
where
\be
C_T(\bfr,\bfr';\beta) = C_0(\bfr,\bfr';\beta) 
\frac{\pi\beta T}{\sin(\pi\beta T)}~,
\label{blochT}
\ee
is the finite-temperature Bloch density matrix, and $\mu$ is the 
chemical potential. In Eq.~(4), the Laplace transform with respect to 
$\mu$ is two-sided, so that $\mu$ is allowed to go negative.  
Specializing now to the case of an isotropic harmonic oscillator in
$d$ dimensions, viz.,
\be
V(\bfr) = \frac{1}{2}m\omega_0^2 r^2,~~~~~~r = \sqrt{x_1^2 + x_2^2 + \cdot
\cdot \cdot+x_d^2}~,
\ee
we have for the zero-temperature Bloch density matrix
\cite{wilson}
\be
C_0^{(d)}(\bfr,\bfr';\beta) = C^{(d)}_0(q,s;\beta) = 
\left(\frac{1}{2\pi}\right)^{d/2}\frac{1}{\sinh^{d/2}(\beta)}
\exp\left\{-\left[q^2\tanh(\beta/2) + (s^2/4)\coth(\beta/2)\right]\right\}~.
\label{bloch0h}
\ee
In the above expression (and what follows), all lengths and energies have been scaled by 
$l_0 = \sqrt{\hbar/m\omega_0}$ and $\hbar\omega_0$, respectively, and we
have introduced the center-of-mass and relative coordinates:
\be
{\bf q} = \frac{1}{2}(\bfr + \bfr'),~~~~~{\bf s} = \bfr - \bfr'~.
\ee

In order to obtain the finite-temperature single-particle density, we
need to set $s=0$ in Eq.~(\ref{bloch0h}) and along with Eq.~(\ref{blochT}), 
perform the inverse Laplace transform
given by Eq.~(\ref{rhoT}).  This is a difficult task if one attempts to 
use Eq.~(\ref{bloch0h})
in its present form.  However, matters can be simplified considerably if one
uses the following identity \cite{brack2,gr},
\be
\exp\{-x\tanh(\beta/2)\} = \sum_{n=0}^{\infty}(-1)^n L_n(2x)e^{-x}
\{e^{-n\beta} + e^{-(n+1)\beta}\}~,
\label{tanh}
\ee
in Eq.~(\ref{bloch0h}), which for $s=0$ now reads
\be
C_0^{(d)}(q;\beta) = \left(\frac{1}{2\pi}\right)^{d/2}\frac{1}{\sinh^{d/2}(\beta)}
\sum_{n=0}^{\infty}(-1)^n L_n(2q^2)e^{-q^2}
\{e^{-n\beta} + e^{-(n+1)\beta}\}~.
\label{uni0}
\ee
Substituting Eq.~(\ref{uni0}) into Eq.~(\ref{blochT}) and performing the 
inverse Laplace transform, Eq.~(\ref{rhoT}),
leads to the finite-temperature single-particle density in any dimension.
In order to illustrate the method, we will now proceed to give an explicit 
calculation for the case of two dimensions (2D), followed by a statement of the
general result in arbitrary dimensions.
%%%%%%%%%%%%%%%%%%%%%%%%%%%%%%%%%%%%
\subsubsection{Two dimensions}
The 2D system turns out to be the simplest case of all,
and for this reason, we give here a detailed derivation of the 
finite-temperature single-particle density.  We begin by noting the
following important exact inverse Laplace transforms (all two-sided):
\be
{\cal L}^{-1}_{\eta}\left[ \frac{e^{-n\beta}}{\sinh(\beta)}\right]
= 2 \sum_{k=0}^{\infty} \delta (\eta - (2k+1) - n)~\Theta(\eta),
\label{twosided}
\ee
\be {\cal L}^{-1}_{\mu} \left[\frac{\pi T}{\sin(\pi\beta T)}\right] =
\frac{1}{\left[\exp(-\frac{\mu}{T})+1\right]}~.
\label{lala}
\ee
Putting $d=2$ in Eq.~(\ref{uni0}) and using Eqs.~(\ref{rhoT}), (\ref{blochT}), 
the finite-temperature density
is given by
\be
\rho^{(2)}(\bfq;T) = \left(\frac{1}{\pi}\right) \sum_{n=0}^{\infty}
(-1)^n L_n(2q^2)e^{-q^2}\times {\cal L}_{\mu}^{-1}\left[
\left(\frac{e^{-n\beta} + e^{-(n+1)\beta}}{\sinh(\beta)}\right)\frac{\pi T}
{\sin(\pi\beta T)}\right]~.
\ee
Applying the convolution theorem for Laplace transforms \cite{pol}
and making use of Eqs.~(\ref{twosided}), (\ref{lala}), we immediately obtain
\bea
\rho^{(2)}(\bfq;T) &=& \left(\frac{2}{\pi}\right) \sum_{n=0}^{\infty}
(-1)^n L_n(2q^2)e^{-q^2}\times 
\sum_{k=0}^{\infty}\left\{ \int_{-\infty}^{\infty} d\tau 
\delta(\tau - (2k+1)-n)\frac{1}{\left[\exp(\frac{\tau-\mu}{T})+1\right]}\right.
\nonumber \\&+& \left. 
\int_{-\infty}^{\infty} d\tau
\delta(\tau - (2k+2)-n)\frac{1}{\left[\exp(\frac{\tau-\mu}{T})+1\right]}
\right\}\nonumber \\
&=& \left(\frac{2}{\pi}\right) \sum_{n=0}^{\infty}
(-1)^n L_n(2q^2)e^{-q^2}\times
\sum_{k=0}^{\infty}\left\{\frac{1}{\left[\exp(\frac{2k+n+1-\mu}{T})+1\right]}
+\frac{1}{\left[\exp(\frac{2k+n+2-\mu}{T})+1\right]}
\right\}\nonumber \\
&=&
\left(\frac{2}{\pi}\right) \sum_{n=0}^{\infty}
(-1)^n L_n(2q^2)e^{-q^2}\times
\sum_{k=0}^{\infty}\frac{1}{\left[\exp(\frac{\varepsilon_n+k-\mu}{T})+1\right]}
\nonumber \\
&=&
\left(\frac{2}{\pi}\right) \sum_{n=0}^{\infty}
(-1)^n L_n(2q^2)e^{-q^2}F^{(2)}_n(\mu)~,
\label{lily}
\eea
where the function $F^{(2)}_n(\mu)$ is defined as
\be
F^{(2)}_n(\mu) = \sum_{k=0}^{\infty}
\frac{1}{\left[\exp(\frac{\varepsilon_n+k-\mu}{T})+1\right]}~,
\label{lucky}
\ee
and $\varepsilon_n = (n+1)$ is the noninteracting energy spectrum 
(in scaled units) of the 2D harmonic
oscillator potential.
In the limit $T\rightarrow 0$, the Fermi function in Eq.~(\ref{lucky}) goes 
over to the
Heaviside step function, and when filling $M+1$ oscillator shells,
Eq.~(\ref{lily}) reduces to
\be
\rho^{(2)}(q) = \left(\frac{2}{\pi}\right) \sum_{n=0}^{M} (M-n+1) 
(-1)^n L_n(2q^2)e^{-q^2}~.
\label{fine}
\ee
Equation (\ref{fine}) is of course identical to the result obtained in 
Ref.~\cite{brack2} in which only zero-temperature quantities were investigated.
\begin{figure}
\resizebox{4in}{5in}{\includegraphics{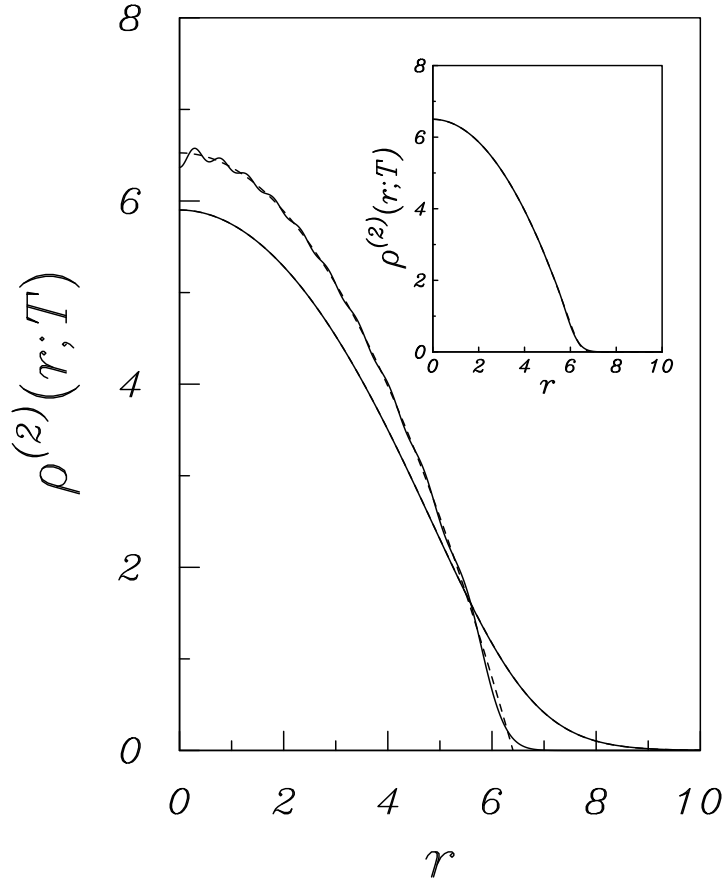}}
\caption{The 2D exact (solid curves) and TF (dashed curves) number densities
at various temperatures for $N=420$ particles.
The upper curves in the main figure correspond
to $T=0$, while the lower-lying curves are for $T=5$.  The figure inset
shows the 2D exact and TF densities at $T=1$.  All quantities have
been scaled as discussed in the text.}
\label{fig1}
\end{figure}
In Fig.~\ref{fig1}, we display the exact 2D 
radial density (solid curves), $\rho^{(2)}(r;T)$, for $N=420$, where 
the lowest $20$ shells 
are filled. For comparison, we have also displayed the corresponding
2D TF densities (dashed curves) at zero \cite{brack2} and 
finite temperatures \cite{note3}.
We have used Eqs.~(\ref{lily}) and (\ref{fine}) for the exact nonzero and 
zero-temperature density respectively. 
In all plots, $T$ denotes the dimensionless quantity 
$({k_B T\over {\hbar\omega}})$.   
We note from Fig.~\ref{fig1} that the oscillations due 
to shell effects are already washed out by $T=1$, 
resulting in TF and exact densities being almost
indistinguishable.

In the high temperature regime where $\mu < \varepsilon_0$, we can make
use of the expansion
\be
\frac{1}{\left[\exp(\frac{2k+\varepsilon_n-\mu}{T})+1\right]} =
\sum_{j=1}^{\infty} (-1)^{j+1}e^{j\mu/T}e^{-j(n+1)/T}e^{-2jk/T}
\ee
in Eq.~(\ref{lily}).   The $k$ sums can be evaluated exactly, and we obtain
\be
\rho^{(2)}(\bfq;T) = 2\sum_{j=1}^{\infty}(-1)^{j+1}
e^{j\mu/T}C^{(2)}_0(\bfq;j/T)~.
\label{fun}
\ee
Making the formal identification of $\beta = 1/T$, 
Eq.~(\ref{fun}) can be re-written as
\be
\rho^{(2)}(\bfq;T) =  2\sum_{j=1}^{\infty}(-1)^{j+1}
e^{j\beta\mu}C^{(2)}_0(\bfq;j\beta)~.
\label{fish}
\ee
A result identical to Eq.~(\ref{fish}) has recently been obtained by Wang
\cite{wang}, but as we have discussed above, this form for the density is
only valid for $\mu < \varepsilon_0$.  In contrast, our Eq.~(\ref{lily}) is 
valid at
all temperatures and appears in a form that clearly displays the role
of the quantum statistics of the system.

Finally, an interesting relationship between $F^{(2)}_n(\mu)$
and the Fermi distribution function can be obtained through
the normalization condition
\bea
N &=& \int d^2q~\rho^{(2)}(\bfq;T)
\nonumber \\ &=&
2 \sum_{n=0}^{\infty}F^{(2)}_n(\mu) \nonumber \\
&=&
2 \sum_{n=0}^{\infty}\frac{(n+1)}
{\left[\exp(\frac{\varepsilon_n-\mu}{T})+1\right]}~.
\eea
%%%%%%%%%%%%%%%%%%%%%%%%%%%%%%%%%%%%%%
\subsubsection{Arbitrary dimensions}
The generalization of the above analysis to arbitrary dimensions
is straightforward.  First, since we are only interested in the
diagonal Bloch density matrix, we can immediately set $s=0$ in Eq.~(\ref{bloch0h}).  Furthermore, we observe
that in any dimension
\be
\sinh^{(1-d/2)}(\beta) = 2^{(d/2-1)}e^{\beta(1-d/2)}\left(
1 + \sum_{m=1}^{\infty}g^{(d)}_m e^{-2m\beta}\right)~.
\label{sinh}
\ee
Multiplying the numerator and denominator of Eq.~(\ref{bloch0h}) by 
Eq.~(\ref{sinh}), and using the identity, Eq.~(\ref{tanh}), we obtain 
\bea
C^{(d)}_{0}(\bfq;\beta) &=&\frac{1}{\pi^{d/2}}\frac{1}{\sinh(\beta)}
\sum_{n=0}^{\infty} (-1)^nL_n(2q^2)e^{-q^2}
\left( e^{-(n+d/2-1)\beta} + e^{-(n+d/2)\beta}\right)
\left(1 + \sum_{m=1}^{\infty} g^{(d)}_m e^{-2m\beta}\right)~.
\eea
Proceeding {\em exactly} as in the 2D case, we finally obtain in
any dimension
\be 
\rho^{(d)}(\bfq;T) = 2\left(\frac{1}{\pi}\right)^{d/2}
\sum_{n=0}^{\infty}F_n^{(d)}(\mu) (-1)^nL_n(2q^2)e^{-q^2}~,
\label{high}
\ee
where
\be
F_n^{(d)}(\mu) = \sum_{k=0}^{\infty}
\left( \frac{1}{\exp[(\varepsilon^{(d)}_n+k-\mu)/T]+1} +
\sum_{m=1}^{\infty}\frac{g_m^{(d)}}{\exp[(\varepsilon^{(d)}_n+k+2m -\mu)/T]+1}
\right)~,
\ee
and $\varepsilon^{(d)}_n = n + d/2$.
The expansion coefficients for $d=1,2,3$ are given by \cite{brack2}
\bea
g_m^{(1)}& =& -(2m-3)!!/(2m)!!~~~~{\rm with}~g_1^{(1)} = -1/2~,\nonumber \\
g_m^{(2)}&=& 0~,\nonumber \\
g_m^{(3)} &=& (2m-1)!!/(2m)!!~.
\eea
One of the most appealing aspects of Eq.~(\ref{high}) is that the function
$F^{(d)}_n(\mu)$ contains all of the dimensional and statistical dependence.
Moreover, for
temperature and particle number ranges where significant deviations from the 
TF approximation are expected (see Fig.~\ref{fig1}), 
the above sums can be truncated
quite quickly.  For example, at low temperature and $N = {\cal O}(10^3)$, 
$(n_{max},k_{max},m_{max}) = (50,50,10)$ is more than adequate to obtain
single-particle density in any dimension. 

Following the 2D example, it is easy to show that in any dimension, the 
high-temperature density (i.e., $\mu < \varepsilon^{(d)}_0$) is given by
\be
\rho^{(d)}(\bfq;T) =  2\sum_{j=1}^{\infty}(-1)^{j+1}
e^{j\beta\mu}C^{(d)}_0(\bfq;j\beta)~.
\ee
In fact, when $\mu < \varepsilon^{(d)}_0$, the following identity for
fermions exists in all dimensions:
\bea
\sum_{j=1}^{\infty}(-1)^{j+1}e^{j\beta\mu}\left(\frac{e^{-j\beta n} + e^{-j\beta(n+1)}}
{[2\sinh(j\beta)]^{d/2}}
\right) = F^{(d)}_n(\mu)~.
\eea
In the $d$-dimensional case, the chemical potential is determined through the 
normalization condition
\bea
N^{(d)} &=& \int d^{(d)}q~\rho(q,T) \nonumber \\
&=& 2\sum_{n=0}^{\infty}(-1)^nF_n^{(d)}(\mu) \times 
F[-n,d/2;1;2]\nonumber \\
&=& 2 \sum_{n=0}^{\infty}\frac{\eta^{(d)}}{\exp[(\varepsilon^{(d)}_n - \mu)/T]+1}~,
\eea
where $F[a,b;c;z]$ is a hypergeometric function \cite{gr}, 
and $\eta^{(d)}$ denotes
the degeneracy of the quantum levels in $d$ dimensions.  The functions
$F[a,b;c;z]$ can only be evaluated analytically for even $d$; their
values are given by $(-1)^nF[-n,d/2;1;2] = 1$, $(2n+1)$, 
$(2n^2+2n+1)$, for $d = 2, 4, 6$, respectively.
%%%%%%%%%%%%%%%%%%%%%%%%%%%%%%%%%%%%%%%%%%%%%%%%%%
\subsection{Single-particle kinetic energy density in $d$ dimensions}
The finite-temperature kinetic energy density can be obtained in a manner 
entirely analogous to the derivation for the single-particle density. 
We begin by first investigating two
expressions at zero temperature (in scaled units)\cite{brack2}
\be
\tau(\bfq) = - 2\frac{1}{2} \sum_{\varepsilon_i<E_f}\phi_i^{\star}(\bfr)
\nabla^2\phi_i(\bfr)~,
\label{kite}
\ee
\be
\tau_1(\bfq) = 2\frac{1}{2} \sum_{\varepsilon_i<E_f}
|\nabla\phi_i(\bfr)|^2~,
\label{kitten}
\ee
where, again, the factor of two accounts for spin.  Equations (\ref{kite}),
(\ref{kitten})
both integrate to the same exact kinetic energy and in the presence of time
reversal symmetry are simply related by
\be \tau(\bfq) = \tau_1(\bfq) - \frac{1}{4}\nabla^2\rho(\bfq)~.\ee
At low temperatures, shell effects give rise to oscillations in $\tau(\bfq)$
and $\tau_1(\bfq)$ that are exactly opposite in phase \cite{rkb1}.  As such,
a convenient quantity to consider is their mean:
\bea
\xi(\bfq) &=& \frac{1}{2}[\tau(\bfq) + \tau_1(\bfq)]\nonumber \\
&=& -\frac{1}{2}\nabla_{s}^{2}\rho(\bfq,{\bf s})|_{s=0}~.
\eea
Starting from Eq.~(\ref{rhoT}) and noting that differentiation commutes with 
the Laplace
inverse operator, we readily obtain after suitable manipulations of hyperbolic
functions the following Laplace inverse:
\be
\xi^{(d)}(\bfq;T) = {d\over 2}{\cal L}_{\mu}^{-1}\left[{C_T^{(d)}
({\bf q};\beta)\over \beta}~\coth(\beta/2)\right]~,
\label{new}
\ee
where $C_T^{(d)}({\bf q};\beta)$ is  defined by Eqs.~(\ref{blochT}, 
\ref{bloch0h}), with $s=0$. This immediately gives 
\be
\xi^{(d)}(\bfq;T) = \frac{d}{(2\pi)^{d/2}}{\cal L}_{\mu}^{-1}\left[
\frac{1}{4\sinh^2(\beta/2)}\sinh^{(1-d/2)}(\beta)e^{-q^2\tanh(\beta/2)}
\frac{\pi T}{\sin(\pi\beta T)}\right]~.
\label{cool}
\ee
In order to proceed, we substitute Eq.~(\ref{sinh}) into (\ref{cool})
and make use of the convolution theorem for (two-sided) Laplace transforms 
just as in
Sec. \ref{fermidensity}.  We require the following exact Laplace inverse
\be
{\cal L}_{\eta}^{-1}\left[\frac{e^{-n\beta}}{4\sinh^2(\beta/2)}\right] =
\sum_{k=0}^{\infty} (k+1) \delta(\eta - (2k+1) - n)~\Theta(\eta).
\ee
A direct evaluation of Eq.~(\ref{cool}) leads to the simple expression for
$\xi(\bfq;T)$, which is valid in any dimension:
\be
\xi^{(d)}(\bfq;T) = \frac{1}{\pi^{d/2}}\frac{d}{2} \sum_{n=0}^{\infty}
G^{(d)}_n(\mu) (-1)^n L_n(2q^2)e^{-q^2}~,
\label{cast}
\ee
where
\be
G_n^{(d)}(\mu) = \sum_{k=1}^{\infty} (2k-1)
\left[
\frac{1}{\exp[(\varepsilon^{(d)}_n + k-1-\mu)/T]+1}
+\sum_{m=1}^{\infty}\frac{g_m^{(d)}}{\exp[(\varepsilon^{(d)}_n + k+2m-1-\mu)/T]+1}
\right]~.
\ee
It is interesting to note that aside from numerical factors, the 
finite-temperature kinetic
energy density and single-particle density 
have identical functional forms in all dimensions.

An alternative, and useful formula for the kinetic energy density, 
$\xi^{(d)}(\bfq;T)$, may be obtained from Eq.~(\ref{new}) by noting that 
\be
\int_q^{\infty}~e^{-q'^2 \tanh (\beta/2)} q' dq' = {1\over 2} \coth (\beta/2)~
e^{-q^2 \tanh (\beta/2)}. 
\ee
It then follows that the kinetic energy density is given by
\be
\xi^{(d)}(\bfq;T)={d\over 2} \int_q^{\infty} \rho^{(d)}({\bf q}';T)~q'dq'~.
\label{knew}
\ee
%%%%%%%%%%%%%%%%%%%%%%%%%%%%%%%%%%%%%%%
\subsection{Reduction to the TF form for $N\rightarrow\infty$}
At low temperatures (i.e., $T=0$)
and small particle numbers, the spatial densities, given
by Eqs.~(\ref{high},36), 
exhibit pronounced shell oscillations, especially in low
dimensions \cite{brack2}.  However, as the number
of particles is increased, these shell effects get washed out and the
spatial densities approach their usual TF form.  In this section we
show rigorously how
the exact $T=0$ density profiles 
reduce to their TF forms as $N\rightarrow \infty$.  For simplicity, we focus
only on the 2D case. 

Consider the exact zero-temperature density given by 
Eq.~(\ref{fine}). Putting $2q^2=t$, and using the 
one-sided Laplace transform 
\be
F(s)=\int_0^{\infty}\rho^{(2)}(t) \exp(-st) dt~,
\ee
it is easy to find that
\be
F(s)={2\over\pi}~\sum_{n=0}^M (M-n+1) 
{({1\over 2}-s)^n\over {({1\over 2}+s)^{n+1}}}~.
\ee
The above $n$ sum may be performed exactly~\cite{note2}, yielding 
\be
F(s)={2\over \pi}~{1\over {8s^2}}\left[\left({1-2s\over {1+2s}}\right)^{M+1}
\{1-2s\}~+~\left( 4sM+6s-1 \right)\right]~.
\label{guess}
\ee
Note that the real part of $s$ must be positive when taking the inverse 
Laplace transform of $F(s)$. 
Hence $\left|{2s-1\over {2s+1}}\right|<1$, 
so that the first term in Eq.~(\ref{guess}) vanishes as $M\rightarrow\infty$. 
This is indeed the case when the number of fermions $N\rightarrow\infty$. 
In this limit, we get
\be
F(s)={M\over {\pi s}}+ \frac{3}{2\pi s}-{1\over {4\pi s^2}}~.
\ee
Taking the inverse Laplace transform with respect to $t$, we immediately 
obtain 
\be
\rho^{(2)}(t)=\left({M\over \pi}+\frac{3}{2\pi}-{t\over {4\pi}}\right)=
       {1\over \pi}\left(M+\frac{3}{2}-{q^2\over 2}\right)~.
\ee
The last expression is the large-N asymptotic TF result for the density 
in $d=2$ dimensions. Its normalization to the correct particle number  
$N = M^2 + 3M + 2$  for $M+1$ filled shells leads to the
Fermi energy  $E_F = M + 3/2 + O(1/M)$  in the large-M limit.

An analogous calculation may be performed for the kinetic energy density  
at $T=0$. Taking $d=2$ in Eq.~(\ref{cast}), and going over to the $T=0$ limit, 
we have 
\be
\xi^{(2)}({\bf{q}};T=0)={1\over \pi}~\sum_{n=0}^M (-1)^n (M-n+1)^2 L_n(t) 
e^{-t/2}~.
\ee
As before, we take the one-sided Laplace transform of the above equation 
with respect to $s$, and perform the $n$ sum exactly. We then take the 
large $M$ limit, Laplace invert back to the $t=2q^2$ variable and  
obtain $(M+3/2-q^2/2)^2/{2\pi}$, which is just the 
TF result for the kinetic energy density.   
%%%%%%%%%%%%%%%%%%%%%%%%%%%%%%%%%%%%%%%%%%%%%%%  
\section{Ideal Bose gas at finite temperature}
A powerful aspect of formalism can be illustrated when
one considers a trapped ideal Bose gas at finite temperature.  In this
case, the finite-temperature Bloch density matrix is simply given by
(compare with Eq.~(5)) 
\be
C_T(\bfq,\bfs;\beta) = - C_0(\bfq,\bfs;\beta)\frac{\pi\beta T}
{\tan(\pi\beta T)}~.
\ee
Noting the following inverse Laplace transform
\be
{\cal L}_{\mu}^{-1}\left[-\frac{\pi T}{\tan(\pi\beta T)}\right] =
\frac{1}{[\exp\left(-\frac{\mu}{T}\right)-1]}~,
\ee
we observe that Bose statistics serve
only to change the behaviour of the functions $F^{(d)}_n(\mu)$
and $G^{(d)}_n(\mu)$.  Explicitly, for the Bose gas we find
\be 
\rho^{(d)}(\bfq;T) = \left(\frac{1}{\pi}\right)^{d/2}
\sum_{n=0}^{\infty}F_n^{(d)}(\mu) (-1)^nL_n(2q^2)e^{-q^2}~,
\label{king1}
\ee
where
\be
F_n^{(d)}(\mu) = \sum_{k=0}^{\infty}
\left( \frac{1}{\exp[(\varepsilon^{(d)}_n+k-\mu)/T]-1} +
\sum_{m=1}^{\infty}\frac{g_m^{(d)}}{\exp[(\varepsilon^{(d)}_n+k+2m -\mu)/T]-1}
\right)~,
\ee
and similarly,
\be
\xi^{(d)}(\bfq;T) = \frac{1}{\pi^{d/2}}\frac{d}{4} \sum_{n=0}^{\infty}
G^{(d)}_n(\mu) (-1)^n L_n(2q^2)e^{-q^2}~,
\label{king2}
\ee
where
\be
G_n^{(d)}(\mu) = \sum_{k=1}^{\infty} (2k-1)
\left[
\frac{1}{\exp[(\varepsilon^{(d)}_n + k-1-\mu)/T]-1}
+\sum_{m=1}^{\infty}\frac{g_m^{(d)}}{\exp[(\varepsilon^{(d)}_n + k+2m-1-\mu)/T]-1}
\right]~.
\ee
Just an in Sec. \ref{fermigas}, all of the dimensional and statistical
dependence is buried in the functions $F_n^{(d)}$ and $G_n^{(d)}$.  
Not surprisingly, the Bose versions of $F_n^{(d)}(\mu)$ and 
$G_n^{(d)}(\mu)$ differ from their Fermi 
counterparts only by the sign in front of unity in the denominator.
\subsubsection{Alternative method}
As it happens, there is an alternative method for evaluating the 
finite-temperature spatial density in the case of Bose statistics which avoids
the summations introduced by the inverse Laplace method.  
One starts by noting that at all temperatures,
the Bose distribution function can be expanded as a power series
\be
\frac{1}{e^{(\varepsilon^{(d)}_n - \mu)/T}-1} = \sum_{j=1}^{\infty}
e^{j\mu/T}e^{-j\varepsilon^{(d)}_n/T}~.
\ee
This immediately allows us to write the finite-temperature density matrix 
in the form (here, the bosons are taken to be spinless)
\be
\rho^{(d)}(\bfr,\bfr';T) = \sum_{j=1}^{\infty}e^{j\mu/T}
\sum_{n=0}^{\infty}\phi^{\star}_n(\bfr')\phi_n(\bfr)
\exp(-j\varepsilon^{(d)}_n/T)~.
\ee
Identifying $\beta = 1/T$ in Eq.~(3) then allows us to write Eq.~(53)
as
\be
\rho^{(d)}(\bfq,\bfs;T) = \sum_{j=1}^{\infty}e^{j\beta\mu}
C_0^{(d)}(\bfq,\bfs;j\beta)~.
\ee
The finite-temperature particle density is obtained by setting 
$s=0$ in Eq.~(54) and the
kinetic energy density is once again evaluated from Eq.~(32).

It is straightforward to show that our expressions for the Bose gas obtained
by the inverse Laplace method lead to results
identical to the so-called alternative method.  
For example,  in the particularly
simple case of two dimensions, we have
\bea
\rho^{(2)}(\bfq;T) &=& \left(\frac{1}{\pi}\right) \sum_{n=0}^{\infty}
(-1)^n L_n(2q^2)e^{-q^2}\times
\sum_{k=0}^{\infty}\left\{\frac{1}{\left[\exp(\frac{2k+n+1-\mu}{T})-1\right]}
+\frac{1}{\left[\exp(\frac{2k+n+2-\mu}{T})-1\right]}
\right\}~.
\eea
Expanding the Bose functions as in Eq.~(52) and performing the $k$ sums exactly,
we obtain ($\beta = 1/T$)
\bea
\rho^{(2)}(\bfq;T) &=& \frac{1}{\pi}\sum_{j=1}^{\infty}\frac{e^{j\beta\mu}}
{2\sinh(j\beta)}
\sum_{n=0}^{\infty}(-1)^n L_n(2q^2)e^{-q^2}
\{e^{-jn\beta} + e^{-j(n+1)\beta}\}\nonumber \\
&=&
\sum_{j=1}^{\infty}e^{j\beta\mu}
C_0(\bfq;j\beta)~,
\eea
by virtue of Eq.~(9).  Indeed, in all dimensions and all temperatures, 
the following identity can be established for bosons
\bea
\sum_{j=1}^{\infty}e^{j\beta\mu}\left(\frac{e^{-j\beta n} + e^{-j\beta(n+1)}}
{[2\sinh(j\beta)]^{d/2}}
\right) = F^{(d)}_n(\mu)~,
\eea
where $F^{(d)}_n(\mu)$ is given by Eq.~(49).  The analogous result for
fermions is given by Eq.~(27). 
%%%%%%%%%%%%%%%%%%%%%%%%%%%%%%%%%%%%%%%
\section{Weakly interacting fermi gas}
Up to this point, we have been considering noninteracting gases in a trap. 
In this 
section we shall investigate (following Ref.~\cite{heiselberg}) a very dilute 
weakly interacting Fermi gas a finite temperature. 
We will consider both three-dimensional (3D) and 2D geometries and examine the
effect of the interaction on the 
degeneracy of levels in a given HO shell. 
%%%%%%%%%%%%%%%%%%%%%%%%%%%%%%%%%%%%
\subsection{Three dimensions} 
We first consider $N$ fermions in a 
3D isotropic harmonic oscillator trap interacting via a contact
pseudo-potential. The Hamiltonian is then given by
\begin{equation}
H=\sum_{i=1}^N \left({{\bf p}_i^2\over 2} + {1\over 2} {\bf r}_i^2\right) + 
4\pi a \sum_{i<j} \delta^3({\bf r}_i-{\bf r}_j)~.
\label{three}
\end{equation}  
In the above, $a$ is the dimensionless s-wave 
scattering length. The mean-field potential at temperature $T$ is 
\begin{figure}
\resizebox{4in}{5in}{\includegraphics{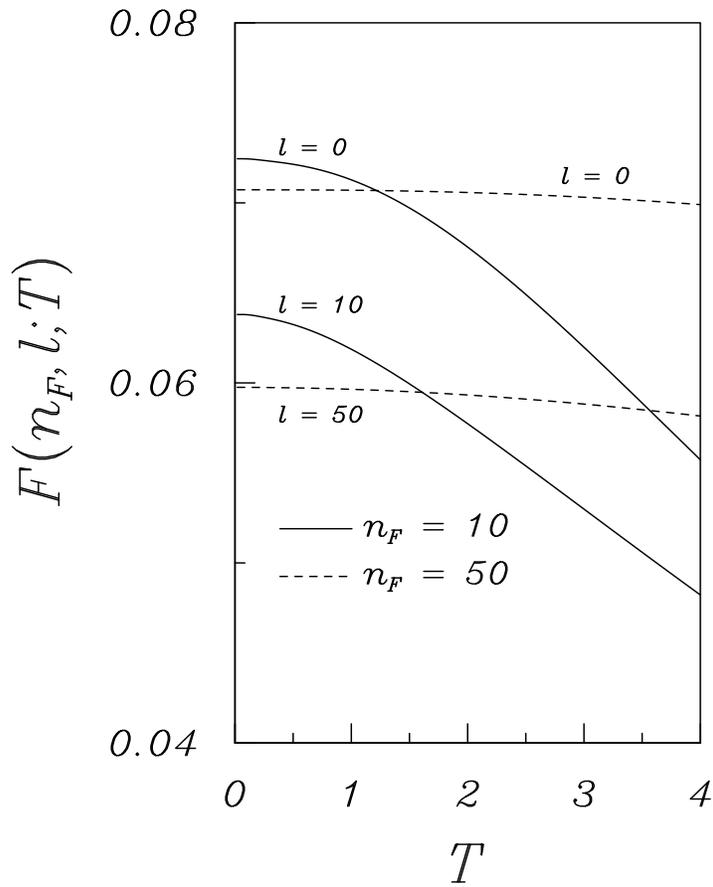}}
\caption{Finite-temperature energy shift, Eq.~(\ref{mot}), for a weakly
interacting 3D trapped Fermi gas as a function of the scaled
temperature $T$.  The solid curves correspond to $n_F=10$ and the
dashed curves to $n_F = 50$.  Note the weak temperature dependence on the
energy shift for $n_F=50$.}
\label{fig2}
\end{figure}
\begin{equation}
U(r;T)=2\pi a \rho(r;T),
\label{int}
\end{equation}
where, keeping in mind that $d=3$, we omit the superscripts on all
quantities. If the 
system is very dilute in the sense that $\hbar\omega \gg |U|$, we may use the 
unperturbed expressions, Eq.~(\ref{high}), for $\rho(r;T)$ to obtain $U(r;T)$. 
To fix the notation for the degenerate states in a shell, we consider the 
noninteracting system at $T=0$, for which the Fermi energy is 
\be
E_F=(n_F+3/2)~,
\ee
and $n_F$ is the principal HO quantum number at the Fermi surface. 
The quantum number $n_F$ is related to the number of fermions, $N$, 
in the well by 
\be
N = {1\over 3} (n_F+1)(n_F+2)(n_F+3)~.
\ee
Note that the shell specified by $n_F$ has degenerate states with 
angular momenta $l=n_F, n_F-2, n_F-4, ...., 1$ or $0$. Consider now a 
particular 
radial HO state ${\cal{R}}_{nl}$, with $(2n+l) = n_F$ in this shell. 
Due to the mean-field potential $U(r;T)$, the single-particle energy of this 
state is shifted (in lowest order perturbation theory) by the amount 
\cite{heiselberg}     
\be
\epsilon_{n_F,l}-E_F=\Delta\epsilon_{n_{F}l}(T)
=\int U(r;T)|{\cal{R}}_{nl}|^2 r^2 dr~.
\label{pert}
\ee
\begin{figure}
\resizebox{4in}{5in}{\includegraphics{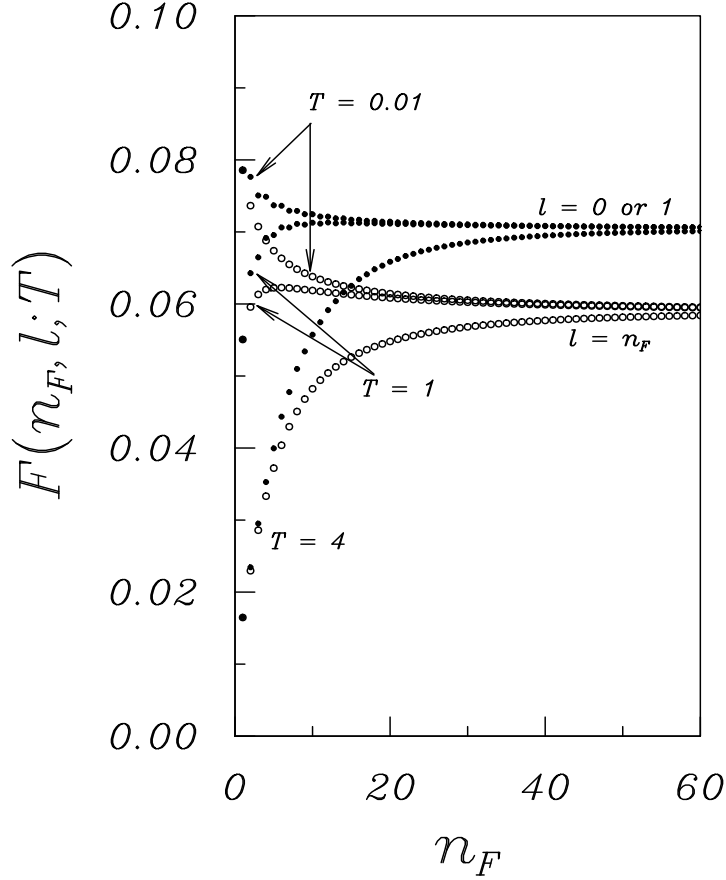}}
\caption{Finite-temperature energy shift, Eq.~(\ref{mot}),
for the 3D weakly interacting
Fermi gas as a function of $n_F$.  
The open circles correspond to $l=n_F$ whereas the filled circles
are for $l=0$ or $l=1$.  
For large  $n_F$, $F(n_F,l;T)$ approaches a constant value, as is
expected in the TF limit.}
\label{fig3}
\end{figure}
Thus the degeneracy of the various angular momentum states is lifted by the 
temperature-dependent interaction (\ref{int}).
Using Eqs.~(\ref{int}) and (\ref{high}), we note that the relevant radial 
integral to obtain the energy shift is of the form (at any temperature $T$)  
\be
I=\int_0^{\infty} r^{2l+2} e^{-2 r^2} L_m(2r^2)\left[L_n^{l+1/2}(r^2)\right]^2 
dr~.
\ee
Although it is possible to obtain an analytical expression for the above 
integral, in practice it is more convenient to calculate it numerically. 
In Fig.~\ref{fig2}, we display the results for the energy shift 
as a function of the scaled temperature $T$ 
for  $l=0$ and $l=n_F $, with $n_F=10$ and $50$.  
Since this energy shift obviously depends on the scattering length $a$ 
and $n_F$, we choose to plot the 
dimensionless quantity 
\be
F(n_F,l;T) = {\Delta\epsilon_{n_{F}l}(T)\over {2\pi 
a N^{1/2} }}~.
\label{mot}
\ee
Our definition of $F(n_F,l;T)$ in Eq.~(\ref{mot}) 
is a little different from that in \cite{heiselberg}, where the
calculation was performed at $T=0$ in the TF approximation. 
Note that the temperature dependence for $n_F=10$ is much more pronounced 
than for $n_F=50$. This would be less so if one scaled the temperature in 
units of the Fermi energy rather than the oscillator spacing. 
We also show in Fig.~\ref{fig3} the level splittings,
$F(n_F,l;T)$, as a function of $n_F$ at various temperatures. 
For large $n_F$, we observe that 
$F(n_F,l;T)$ approaches a constant, demonstrating that 
$\Delta\epsilon_{n_{F}l}(T)$ behaves as 
$N^{1/2}\simeq n_F^{3/2}$, as is to be expected in the TF limit.
%%%%%%%%%%%%%%%%%%%%%%%%%%%%%%%%%%
\subsection{Two dimensions}
\begin{figure}
\resizebox{4in}{5in}{\includegraphics{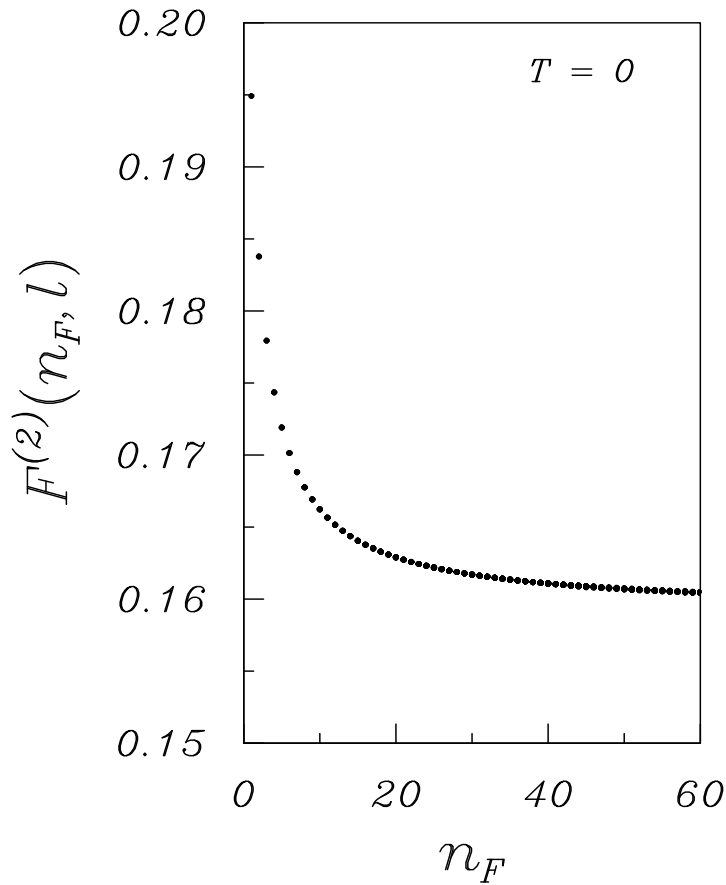}}
\caption{The energy splitting $F^{(2)}(n_F,l)$ defined by Eq.~(71)
for a 2D Fermi gas at $T=0$.  Note that $F^{(2)}(n_F,l)$ is
{\em independent} of $l$ for every HO shell.}
\label{fig4}
\end{figure}

We now calculate the corresponding energy shift in the quasi-2D
harmonic oscillator trap. A highly anisotropic 3D oscillator 
will behave as such when the oscillator frequency in one direction (say 
$\omega_z$) is much, much greater than in the other two ($\omega_x=\omega_y=
\omega_\perp)$. 
Furthermore, the temperature $T$ under consideration is low enough so 
that $k_BT\ll \hbar\omega_z$. If the interatomic interaction is so weak that 
it cannot cause excitations in the $z$ direction, we may restrict the 
Hilbert space in the $z$ direction by setting the oscillator quantum number 
$n_z=0$. The effective interaction of the quasi-2D system is 
then obtained by taking the 3D delta-function potential in 
Eq.~(\ref{three}), and taking its expectation value with respect to the 
HO wave function in the $z$ direction, 
\begin{equation}
\phi_0(z)=\left({\omega_z\over \pi}\right)^{1/4}~\exp(-\omega_z z^2/2)~.
\end{equation}  
The resulting quasi-2D Hamiltonian is then given by (again, scaled units
are used)
\begin{equation}
H=\sum_{i=1}^N \left({{\bf p}_i^2\over 2} + {1\over 2} {\bf r}_i^2\right) + 
g\sum_{i<j} \delta^2({\bf r}_i-{\bf r}_j)~,
\label{two}
\end{equation}
where the momenta and coordinates are planar vectors, $g=2\sqrt{2\pi}(a/\ell_z)$
is the effective 2D dimensionless coupling constant, and we have denoted the 
oscillator length in the $z$ direction by $\ell_z$. 
We restrict this calculation to $T=0$, although the extension to
finite temperature is straightforward. The mean-field one-body potential is 
then given by
\begin{equation}
U^{(2)}(r)=\frac{g}{2}\rho^{(2)}(r)~,
\end{equation}
where the 2D $\rho^{(2)}(r)$ is given by Eq.~(\ref{fine}).
Identifying $M$ with $n_F$, we note that for the 2D case, 
\begin{equation}
N=(n_F+1)(n_F+2)~,~~~~E_F=(n_F+1)~.
\end{equation}   
The radial 2D HO wavefunction is given by 
\begin{equation}
R_{nl}(r)=\left[{2 n!\over {(n+|l|)!}}\right]^{1/2} r^{|l|}e^{-r^2/2}
L_n^{|l|}(r^2)~,
\end{equation}
with the normalization $\int_0^{\infty} dr r |R_{nl}|^2=1$.
The energy shift analogous to Eq.~(\ref{pert}) is 
\begin{equation}
\Delta\epsilon^{(2)}_{n_F,l}=\int_0^{\infty} dr r U(r) |R_{nl}|^2~=\sqrt{2\pi} 
\left({a\over \ell_z}\right)\int_0^{\infty}dr r \rho(r) |R_{nl}|^2~,
\label{pert2}
\end{equation}
and
\begin{equation}
F^{(2)}(n_F,l)\equiv{\Delta\epsilon^{(2)}_{n_F,l}\over {\sqrt{2\pi}({a\over \ell_z}) N^{1/2}}}~.
\label{2star}
\end{equation}
The 2D energy shift at $T=0$ is shown in Fig.~\ref{fig4}.
It readily seen from this figure that $F^{(2)}(n_F,l)$ depends on 
$n_F$, but not explicitly on $l$.
Consequently, unlike the 3D case, there is no splitting (in the 
lowest order perturbation theory)
between states with different $l$ values in a given shell at $T=0$. This 
surprising result may  be also checked  analytically.  
Specifically, a little
algebra gives
\begin{equation}
\Delta\epsilon_{n_F,l}^{(2)}=2\sqrt{{2\over \pi}}~g
{n!\over {(n+|l|)!}} 
\sum_{m=0}^{n_F} (-1)^m (n_F-m+1)\int_0^{\infty}dx x^{|l|} e^{-2x}L_m(2x)
\left[L_n^{|l|}(x)\right]^2~.
\end{equation}
The integral above can be performed exactly by noting that~\cite{gr}
 \begin{equation}
\left[L_n^{|l|}(x)\right]^2={(n+|l|)!\over {n!}}\sum_{k=0}^n 
{x^{2k}\over {(|l|+k)! k!}} L_{n-k}^{|l|+2k}(2x)~.
 \end{equation} 
The final result is given by (with the restrictions 
$(n-k)\leq m\leq (n+k+|l|)$)
\begin{equation}
\Delta\epsilon_{n_F,l}^{(2)}=2\sqrt{{1\over {2\pi}}}~g
\sum_{m=0}^{n_F} (n_F-m+1) 
\sum_{k=0}^n {(-1)^{n-k}\over {2^{|l|+2k}}}~{(|l|+2k)! (n+k+|l|)!\over 
{k!(n-k)!(|l|+k)!(m-n+k)!(n+k+|l|-m)!}}~.
\end{equation}
An explicit case by case evaluation of the above expression~\cite{note2} 
verifies that the shift depends only on $n_F$, and 
is independent of $|l|$; we have not been able to demonstrate this fact
analytically for a general $n_F$.
As stated earlier, the independence of $\Delta\epsilon_{n_F,l}^{(2)}$
on $|l|$ is to be expected for 
large $n_F$ in which the TF density depends only on $r^2$. 
What is surprising is 
that the shift is independent of $l$ for all $l$ within a given shell even
though the {\em exact} single-particle density, Eq.~(16), has been used.
%%%%%%%%%%%%%%%%%%%%%%%%%%%%%
\section{Summary and Discussion}
We have presented for the first time
exact analytical expressions for the number and 
kinetic energy densities of an ideal Fermi gas in a harmonic 
trap at finite temperatures.  
These new expressions, given by Eq.~(\ref{high}) and (\ref{cast}),
are valid at any temperature, and in any dimension.  
We have also derived the corresponding results
for bosons, which are given by Eqs.~(\ref{king1}) and (\ref{king2}).
Unlike previous expressions in the literature (see e.g., Ref.~\cite{wang}),
our functional forms for the particle and kinetic energy densities clearly 
illustrate the role of the quantum statistics 
and have a universal form independent of the dimension of the system. 
We have also demonstrated analytically in Sect.~II C (for $d=2$), how the exact 
fermionic expressions for the particle and kinetic energy densities 
at $T=0$ reduce to 
their TF forms as $N\rightarrow\infty$.   It is worthwhile noting that 
establishing
this reduction analytically using a wavefunction representation for the
particle and kinetic energy densities is a non-trivial task
(see e.g., Ref.~\cite{gleisberg}).

To complement our ideal gas results, we have also considered a weakly 
interacting trapped Fermi gas.
The exact fermionic expression for the spatial density was used to obtain the 
mean-field 
potential, and the finite temperature
first-order shifts in the energy levels for a given HO shell. 
Similar calculations were
recently reported by Heiselberg and Mottelson~\cite{heiselberg} in
3D, but only at $T=0$ and within the TF approximation. 
As was pointed out in Ref.~\cite{heiselberg}, for attractive interparticle 
interactions (i.e., negative $a$), there are also 
pairing correlations between time-reversed states $(l,m)$ and $(l,-m)$ in 
a given shell. In the situation when the splittings given by 
Eq.~(\ref{pert}) are small compared with the pairing gap, the pairing may 
take place between all states of a shell leading to supergaps. 
We plan on performing a detailed study of the supergap phenomenon at 
finite temperatures in a future publication.

As our final application, we also studied the energy level splittings in a 
$d=2$ isotropic HO geometry at $T=0$.   We found the surprising result that
the $l-$degeneracy for a given HO shell {\em is 
not} lifted in the lowest order perturbation theory, even when the exact 
expression for the single-particle density is used. 
In the 2D TF approximation, this result can be understood by noting that the
TF particle density, $\rho^{(2)}_{\rm TF}(r)\propto (\mu - r^2/2)$,
does not break the $SU(2)$ symmetry of the harmonic trap.
However, the radial 
dependence for the exact density (especially for a small number of fermions)  
is more complicated, and therefore our result is unexpected. 
Given this finding, it should prove very interesting to study
the nature of pairing transitions in a 2D trapped Fermi system at
finite temperature. 
%%%%%%%%%%%%%%%%%%%%%%%%%%%%%%%%%%%%%%%%%
\begin{acknowledgments}
This work was supported in part by the National Sciences and the Engineering
Research Council of Canada (NSERC) and by the Canadian Institute for Advanced
Research (CIAR).  
MB acknowledges travel support by the Deutsche Forschungsgemeinschaft.
BVZ would like to thank Dr.~J.~P.~Carbotte for additional financial support.
\end{acknowledgments} 
\bibliographystyle{pra}

\begin{thebibliography}{20} 
\bibitem{bec} F. Dalfovo, S. Giorgini, L. P. Pitaevskii, and S. Stringari, 
Rev. Mod. Phys. {\bf 71}, 463 (1999).
\bibitem{jin} B. DeMarco and D. S. Jin, Science {\bf 285}, 1703 (1999); 
M. J. Holland, B. DeMarco, and D. S. Jin, Phys. Rev. A {\bf 61}, 
053610 (2000). 
\bibitem{brack2} M. Brack and B. P. van Zyl, Phys. Rev. Lett. {\bf 86},
1574 (2001).
\bibitem{schneider} J. Schneider and H. Wallis, Phys. Rev. A {\bf 57}, 1253
(1998).
\bibitem{brosens} F. Brosens, J. T. Devreese, and L. F. Lemmens, Phys. Rev. E
{\bf 57}, 3871 (1998).
\bibitem{vig} P. Vignolo, A. Minguzzi, and M. P. Tosi, Phys. Rev. Lett. 
{\bf 85}, 2850 (2000).
\bibitem{akdeniz} Z. Akdeniz, P. Vignolo, A. Minguzzi, and M. P. Tosi, 
cond-mat/0205480 v1 (2002).
\bibitem{bruun} G. M. Bruun and C. W. Clark, Phys. Rev. A {\bf 61}, 061601
(2000).
\bibitem{gleisberg} F. Gleisberg, W. Wonneberger, U. Schl\"oder, and
C. Zimmermann, Phys. Rev. A {\bf 62}, 063602 (2000).
\bibitem{wang} X. Z. Wang, Phys. Rev. A {\bf 65}, 045601 (2002).
\bibitem{butts} D. A. Butts and D. S. Rokhsar, Phys. Rev. A {\bf 55},  
4346 (1997).  
\bibitem{note1} Specifically, the results for fermions in Ref.~\cite{wang} are
only valid for temperatures where $\mu < \varepsilon_0$ (see Sec.~IIA).
\bibitem{lieb} E. H. Lieb, Rev. Mod. Phys. {\bf 53}, 603 (1981); 
{\bf 54}, 311 (1982) {\it erratum}.
\bibitem{heiselberg} H. Heiselberg and B. Mottelson, Phys. Rev. Lett. 
{\bf 88},190401 (2002).
\bibitem{pol} B. van der Pol and H. Bremmer, {\it Operational Calculus}, 
(Cambridge University Press, Cambridge, U.K.), Second Ed. (1955).
\bibitem{brack} M. Brack and R. K. Bhaduri, {\em Semiclassical Physics},
Addison-Wesley, part of the Frontiers in Physics vol.~96 (1997).
\bibitem{wilson}E. H. Sondheimer and A. H. Wilson, Proc. R. Soc. London
A {\bf 210}, 173 (1951).
\bibitem{gr} I. S. Gradshteyn and I. M. Ryzhik, {\em Table of Integrals,
Series, and Products} (Academic Press, New York, 1994), 4th ed.,
Eq.~8.975.1. Also see 
Eq.~(8.976.4). Note that there is a misprint in the latter formula; the upper 
limit in the sum is not infinity, but $n$.
\bibitem{note3} In two dimensions, the finite-temperature TF density is given
by $\rho^{(2)}_{\rm TF}(r;T) = (T/\pi)\ln(1 + \exp[\beta(\mu-V(r))])$.
\bibitem{rkb1} R. K. Bhaduri and L. F. Zaifman, Can. J. Phys. {\bf 57},
1990 (1979); C. Guet and M. Brack, Z. Phys. A {\bf 297}, 247 (1980).
\bibitem{note2} We have used the mathematical software package,  
Maple$^\copyright$, to evaluate this quantity.  


\end{thebibliography}

\end{document}